\documentclass[twoside,twocolumn,prc,floats,showkeys,amsmath,amssymb]{revtex4-2}

\usepackage{xcolor}

\usepackage{graphicx}
\usepackage{textcomp,nicefrac}
\usepackage{tabularx}
\usepackage[normalem]{ulem}
\usepackage{enumitem}

\begin{document}

\title{A Data-Driven Reconstruction Technique based on Newton's Method for Emission Tomography  }
\author{Loizos Koutsantonis}%
\email{loizos.koutsantonis@uni.lu}
\author{Tiago Carneiro}
\email{tiago.carneiropessoa@uni.lu}
\author{Emmanuel Kieffer}
\email{emmanuel.kieffer@uni.lu}
\author{Frederic Pinel}
\email{frederic.pinel@uni.lu}
\author{Pascal Bouvry}
\email{pascal.bouvry@uni.lu}

\affiliation{Department of Computer Science, Faculty of Science, Technology and Medicine, University of Luxembourg, Luxembourg}

\date{\today}
\begin{abstract}
\textbf{\abstractname:} In this work, we present the Deep Newton Reconstruction Network (DNR-Net), a hybrid data-driven reconstruction technique for emission tomography inspired by Newton's method, a well-known iterative optimization algorithm. The DNR-Net employs prior information about the tomographic problem provided by the projection operator while utilizing deep learning approaches to a) imitate Newton's method by approximating the Newton descent direction and b) provide data-driven regularisation.
 We demonstrate that DNR-Net is capable of providing high-quality image reconstructions using data from SPECT phantom simulations by applying it to reconstruct images from noisy sinograms, each one containing 24 projections. The Structural Similarity Index (SSIM) and the Contrast-to-Noise ratio (CNR) were used to quantify the image quality. We also compare our results to those obtained by the OSEM method. According to the quantitative results, the DNR-Net produces reconstructions comparable to the ones produced by OSEM while featuring higher contrast and less noise. 
\end{abstract}
\maketitle

%\IEEEpeerreviewmaketitle

\section{Introduction}

Positron Emission Tomography (PET) and Single Photon Emission Computed Tomography (SPECT) are the two most used modalities in nuclear medicine. Utilizing radioactive tracers, PET and SPECT allows the in-vivo tomographic imaging of the physiological function of the tissue or organ being examined. Both modalities use sophisticated image reconstruction algorithms to transform the projections of the radioactivity distribution obtained at different angles into tomographic images. However, due to the limited amount of the radioactive tracer that can be administrated to the patient, the projection data in PET and SPECT usually result in noisy and low-quality reconstructed tomographic images \cite{Garcia, Wie, WangX}. 

The diagnostic capabilities of PET and SPECT are strongly depended on the quality of the produced tomographic images. Applications of nuclear medicine, including the detection of defects in cardiac imaging, the localization of tumors at an early stage, and the assessment of patients with Parkinson's disease, require images of sufficiently high Contrast-to-Noise Ratio (CNR) and high resolution.  Hence, improving image quality while maintaining the radiation dose at its minimum possible level is a challenging task in nuclear medicine \cite{WangX, Ouyang, Lim}.

Various physical model-based algorithms have been proposed in the last two decades to provide accurate tomographic reconstructions from low-count PET, and SPECT sinograms \cite{Fessler, He, Nuyts, Wolf, WangC, Boud}. These techniques use a forward imaging model linking the tomographic image to the sinogram and incorporate prior knowledge on the solution to penalize the image variance or, equivalently, increase the image CNR.

Many alternative techniques apply Artificial Intelligence (AI),  particularly Deep Learning (DL), to address the challenging task of tomographic image reconstruction from a set of noisy projections \cite{Reader, Hag, Zhu, Feng, Shao, Lim, Chrys, Whitel}. These techniques utilize a deep neural network built and trained to provide the direct mapping between the projection space and image space. After proper training with noisy samples, the model is able to directly transform a low-count sinogram into a  tomographic image of sufficiently good quality.

The majority of the AI-based approaches do not exploit any physical model providing information about the PET/SPECT tomographic problem \cite{Lim}. Instead, their success relies on the deep neural network's capability to approximate the inverse of the attenuated Radon transform and produce image solutions of the desired quality and smoothness. In general, these approaches are required to learn the geometric properties of the tomographic problem and the physical factors, such as the photon attenuation and scatter, affecting the quality of the obtained planar projections. Thus, a large set of samples is needed for training the DL model to accurately describe the inverse mapping between the sinogram and the image. 

\begin{figure*}[t!]
\centering
\includegraphics[width = \textwidth]{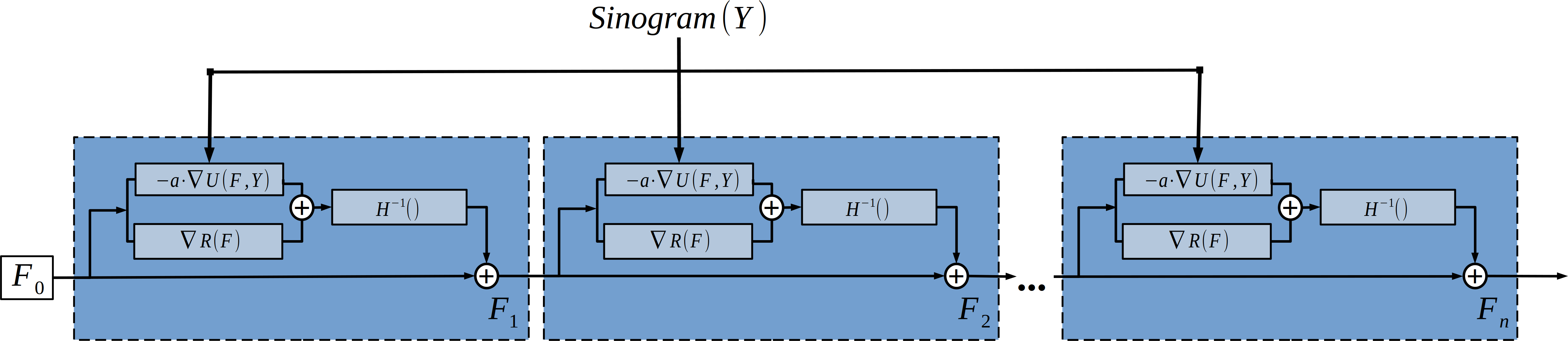}
\caption{The Deep Newton Reconstruction Network (DNR-Net) proposed for image reconstruction in emission tomography. The DNR-Net unrolls the  Newton iterative formula into a sequence of blocks corresponding to a single Newton iteration. The network maps the sinogram $Y$ to a tomographic image $F_n$ by successively updating the image $F$ using the operator  $H^{-1}(\nabla U(F_i|Y)+ \nabla R(F_i))$. This operator is built on a linear component $-\nabla U(F_i|Y)$ incorporating prion information on the tomographic problem  (see Eq. \ref{Eq:Nabla}). The other two components, $\nabla R()$ and $H^{-1}()$, are two neural networks trained to regularize the reconstructed image and approximate the Newton descent direction, respectively.  The architecture of a single block is shown in Fig. \ref{fig:model}.}
\label{fig:arch}
\end{figure*}

In this work, we present the Deep Newton Reconstruction Network (DNR-Net), an end-to-end trained data-driven method for image reconstruction in SPECT and PET. Inspired by Newton's method, DNR-Net provides a hybrid image reconstruction method that incorporates the physical model of the tomographic problem as encoded by the system matrix within the DL architecture. This hybrid approach produces reconstruction results using a training set of reduced size, compared to  pure DL-based methods proposed for tomographic image reconstruction  in different studies \cite{Hag, Hu, Chrys, Shao1}.

\section{The  Image Reconstruction Problem}

The tomographic image reconstruction in PET and SPECT is an inverse problem characterized by noisy and attenuated data. Moreover, especially in the case of SPECT, the planar projections obtained about the body are limited in number, making this problem ill-posed.  The traditional iterative  methods with proven efficacy in SPECT/PET image reconstruction formulate this inverse problem as a  optimization problem consisting of a data fidelity term $U(F)$ quantifying the linkage between  the image $F = \{f_1, f_2,..,f_n\} $ and the sinogram $Y = \{y_1, y_2,..,y_m\}$,  and a regularisation term $R(F)$  inducing sparsity constraints on the image solution $\hat F$ \cite{Fessler}:
\begin{equation}
     \hat F = \arg \min_ {F} U(F|Y) +  R(F)
     \label{Eq:Cost}
\end{equation}
For SPECT and PET, the term $U(F)$ is the negative Poisson log-likelihood function measuring the "goodness" of the reconstructed image $F = \{f_1, f_2,..,f_n\} $ in representing  the measured sinogram counts $Y = \{y_1, y_2,..,y_m\}$:
\begin{equation}
U = \sum_i^m\bigg(\tilde{y_i} - y_i\ln \tilde{y_i} + \ln y_i! \bigg)
\label{Eq:Loglikelihood}
\end{equation}
where $\tilde{y_i}$  are the  expected sinogram counts estimated from the image pixels $f_j$ using the system matrix  $\mathbf{A}$:  
\begin{equation}
    \tilde{y_i} = \sum_j^n\mathbf{A_{ij}}f_j
    \label{Eq:Proj}
\end{equation}
In the above parametrization of the problem,  the system matrix $\mathbf{A}$ acts as a forward-operator; it provides a probabilistic model for the emission and attenuation of $\gamma$-photons before exiting the body and encodes the detector position and geometry during the data acquisition. 

\section{The Deep Newton Reconstruction  Network }

\subsection{Model Architecture}

The classical Newton's method is a second-order iterative algorithm that can be used for solving optimization problems such as that of Eq. \ref{Eq:Cost}. Compared to first-order algorithms such as the well-known gradient descent, Newton's method exploring locally the curvature of the optimization function has superior convergence properties. 
 Given the optimization function $J$ defined on the parameters vector $\Vec{x}$, Newton's iterative formula is given by \cite{Arfken}: 
\begin{equation}
    \Vec{x}_{i+1} = \Vec{x}_i - H^{-1}\nabla J(\Vec{x}_i)
    \label{Eq:Newton}
\end{equation}
where, $H^{-1}$ is the inverse of the Hessian matrix, $H = \nabla^2 J(\Vec{x}_i)$, providing second-order derivative information on the curvature of the optimization function $J$ at $\Vec{x}_i$. For a vector $\Vec{x}$  of size $N$,  the inversion of the Hessian matrix having a size of  $N \times N$ can be performed using a linear solver with a computational complexity of  $\mathcal{O}(N^3)$ \cite{Syamal}. However, the vast computational cost required for the inversion of the Hessian in large-scale problems makes the application of Newton's method impractical \cite{Haber}.   Variants of Newton's method using inexact approximations of the Hessian matrix have been used in large-scale optimization and inverse problems \cite{Haber, Roosta, Zey, HuJ, WangH, Tsai} to provide faster convergence. Such a case is the problem of emission tomography where the typical dimension $N$ of the image vector $\Vec{x} = F$ is $128 \times 128$.

\begin{figure*}[t!]
\centering
\includegraphics[width = 2\columnwidth]{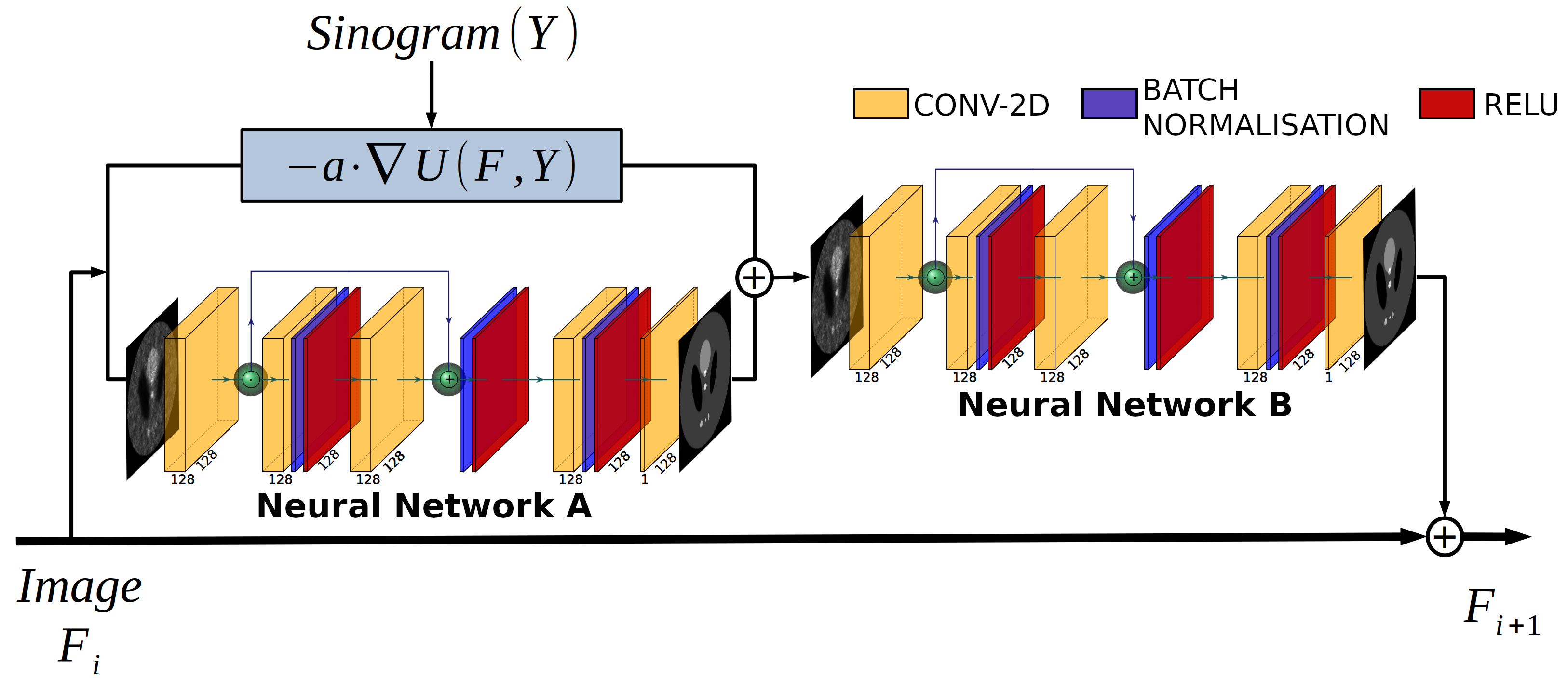}
\caption{The unit block of the DNR-Net. A residual network is used for building each of the two trained operators in each block ("Neural Network A"/"B"), which, in this study, had identical architectures. DNR-Net was configured by six sequential blocks  end-to-end trained using synthetic SPECT sinograms generated from numerical phantoms. }
\label{fig:model}
\end{figure*}

Inspired by Newton's method, we propose the Deep Newton Reconstruction Network (DNR-Net) for approximating the inverse of the Hessian matrix in solving the problem of emission tomography. DNR-Net is a hybrid method developed to overcome the limitation of the computationally demanding inversion of the Hessian matrix; it uses prior information provided by the system matrix and employs deep learning techniques to approximate the Newton descent direction directly.

For the optimization problem presented in Eq. \ref{Eq:Cost}, the Newton iterative formula takes the form:
\begin{equation}
   F_{i+1} = F_i - H^{-1}(\nabla U(F_i|Y)+ \nabla R(F_i)) 
   \label{Eq:NT}
\end{equation}

In DNR-Net, the iterative formula of Eq. \ref{Eq:NT} is unrolled into a sequence of $n$ blocks composing a deep architecture as shown in Fig. \ref{fig:arch}.  Each block corresponding to a single Newton iteration acts on the image produced by its previous block. The sinogram is used as an input to all blocks configuring this architecture.  The unit block of DNR-Net (depicted in Fig. \ref{fig:model}) comprises of the three operators, which are described below:
\begin{enumerate}[label=\Alph*.]
\item \textbf{Linear Operator $-a\nabla U(\cdot)$}: A differential operator acting on the output image  from the previous block $F_i = \{f_1, f_2,..,f_n\} $. It calculates the partial derivatives of the negative log-likelihood function in Eq. \ref{Eq:Loglikelihood} using the sinogram measurements $Y = \{y_1, y_2,..,y_m\}$ and the system  matrix $\mathbf{A_{ij}}$ of the  tomographic problem:

\begin{equation}
    \nabla U_j = \frac{\partial{U}}{\partial{f_j}} = \sum_i^m \bigg( \mathbf{A_{ij}}- y_i\frac{\mathbf{A_{ij}}}{\mathbf{\sum_j^n A_{ij}} f_j} \bigg)
    \label{Eq:Nabla}
\end{equation}
\item \textbf{Regulariser $\sim {\nabla R}(\cdot)$}: A deep neural network replacing the regularisation term in Eq. \ref{Eq:Cost}. This neural network learns to regularise the image produced from the previous block, which is fed at its input.  Its output is added to the linear operator output, and the result is fed to the Newton Direction Estimator. The regulariser is shown as "Neural Network A" in the architecture of Fig. \ref{fig:model}.
\item \textbf{Newton Direction Estimator $\sim H^{-1}\nabla J(\cdot)$}: A deep neural network trained to approximate the Newton descent direction using the resulting output from the two previous operators. The  output from this operator is used to update the input image: $F_{i+1} = F_i + H^{-1}\nabla J(\cdot)$. The result is forwarded to the input of the next block. The Newton direction estimator is shown as  "Neural Network B" in the architecture of Fig. \ref{fig:model}.
\end{enumerate}

\subsection{Model Configuration} 

The number of blocks in DNR-Net is a hyperparameter arbitrarily chosen. In this study, DNR-Net is built by 6 unit blocks having the same architecture.  The two trained operators of each unit block were implemented as two identical Residual Networks (ResNet) \cite{Kaiming}. Each ResNet comprises two residual blocks, each consisting of two convolution layers followed by a leaky Rectified Linear Activation Unit (ReLU).  A batch normalization layer was placed between the second convolution layer and the non-linear activation function of each residual block. A single residual block is depicted in the schematics of "Neural Network A" and "B" in Fig. \ref{fig:model}. 

DNR-Net is configured to reconstruct $128 \times 128$ images at its output from input sinograms containing 24 projections.  The result from the application of the linear operator (Eq. \ref{Eq:Nabla}) on an image of uniform activity distribution is used to initialize the input image $F_0$ of the first unit block in DNR-Net. 

\subsection{Model Training}
\label{Sec:Training}
\begin{figure}[t!]
\centering
\includegraphics[width = \columnwidth]{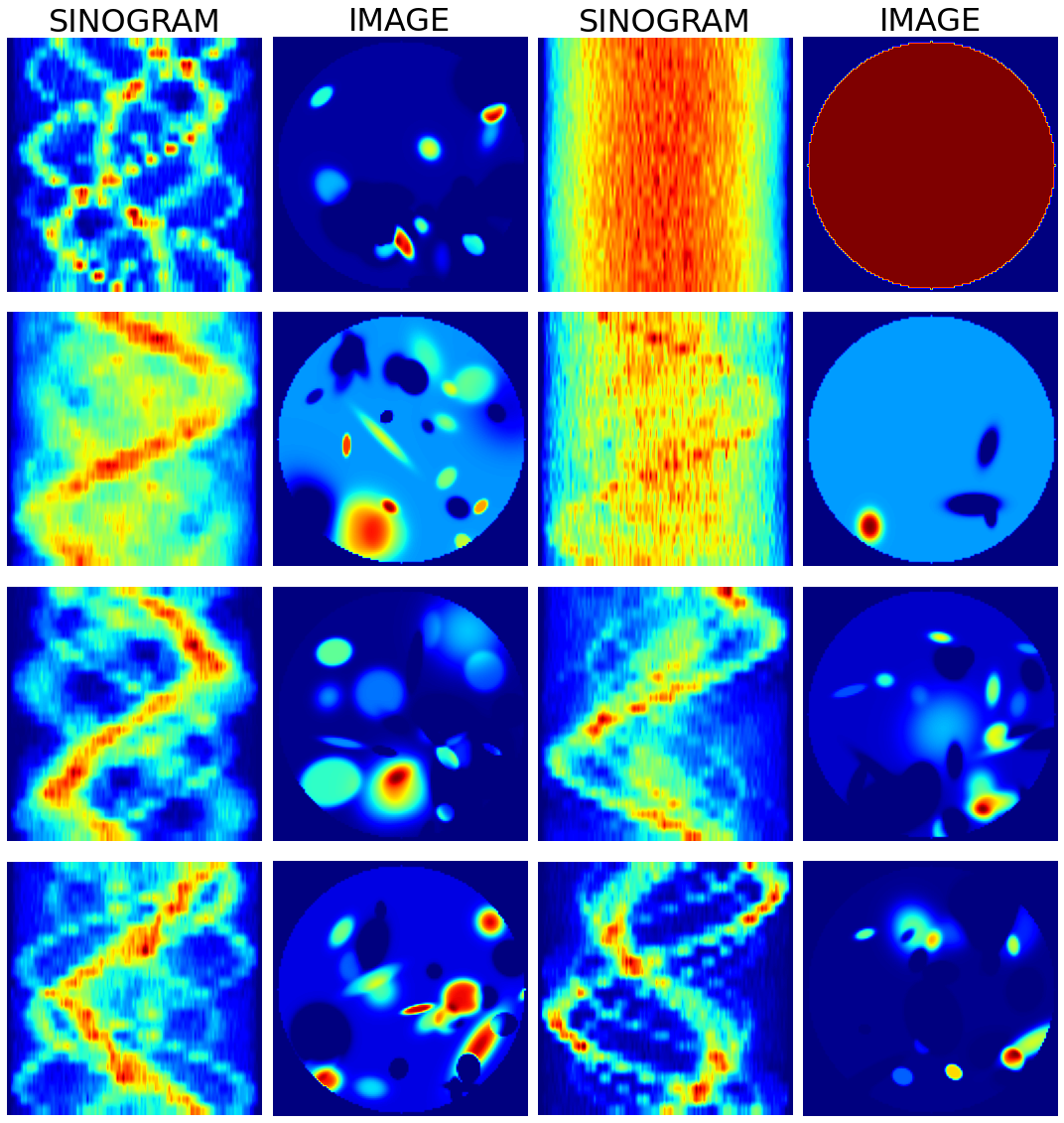}
\caption{A total of 3000 images and corresponding SPECT sinograms were used for training the DNR-Net. Image and sinogram samples are shown in the above figure.  The sinograms were simulated from randomly defined images using the exact forward model of Eq. \ref{Eq:Proj} and further randomized with Poisson noise. }
\label{fig:samples}
\end{figure}

The  Mean Squared Error (MSE) was used as the loss function for the training of the  DNR-Net. MSE is given by:
\begin{equation}
    L(F^{m}, F^{0}) = \frac{1}{N^2}\sum_{i=1}^{N^2}(F^{m}_i - F^{0}_i)^2
\end{equation}
where, $F^{0}$ and $F^{m}$ are the ground truth and predicted by the DNR-Net images respectively, and $N^2$ is the image size. 
The training set of  ground truth images was generated using the "Fermi-like" model introduced  in \cite{Koutsantonis, Koutsantonis1} to train the proposed architecture. This model uses a superposition of $K$ ellipsoidal sources  to parametrize the tomographic image of the activity distribution: 
\begin{equation}
    F(x_i,y_j) = A_0 + \sum_{k=1}^K  A_k \bigg( \exp{\frac{r_k(x_{ck}, y_{ck}) - R_k}{d_k R_k} +1} \bigg)^{-1}
\end{equation}
where $(x_i,y_j)$ are the physical coordinates of the image pixels, $A_0$ is a uniform background coefficient, $A_k$ is the activity amplitude of the $k^{th}$ source, $r_k$ is the euclidean distance of the $ij^{th}$ pixel from the $k^{th}$ source center $(x_{ck}, y_{ck})$, $d_k$ is the diffusion constant of the $k^{th}$ source, and $R_k$ is the geometrical factor: 
\begin{equation}
R_k  = \frac{uv}{\sqrt{u^2 cos^2(\theta - \phi) + v^2 sin^2(\theta - \phi)}}
\end{equation}
defined by the major- and minor- axes $(u, v)$, and orientation angle $\phi$,  of the ellipsoidal source in the tomographic plane. $\theta$ is the angle between the major axis $u$ of the source and the line segment connecting the source center to the $ij^{th}$ pixel. All parameters in the above model, including the number of sources $K$ were randomized in defined limits to generate a training set of 3000  ground truth tomographic images. 
SPECT sinograms were simulated from the generated images using the exact forward model of Eq. \ref{Eq:Proj} and randomized with Poisson noise. The sinograms were simulated for 24 views in the $360^{\circ}$ angular range assuming an equiangular parallel projection geometry.  Representative image samples and corresponding sinograms used in the training set are shown in Fig. \ref{fig:samples}. 

\subsubsection*{Aspects of implementation}
The DNR-Net architecture was implemented using the  Pytorch 1.7.1 framework in Python 3.8.  The computationally intensive linear operator of Eq. \ref{Eq:Nabla} is programmed in C using the Compute Unified Device Architecture (CUDA-C) to exploit the massively parallel hardware of Graphics Processing Units (GPUs). The pre-compiled CUDA kernel is integrated into the PyTorch implementation of DNR-Net by using an interface mechanism. In turn, the CUDA version used in the training process is 11.2. 

DNR-Net was end-to-end trained on a testbed equipped with a 2.80 GHz Intel Xeon W-10855M CPU and 16 GB RAM. The GPU used in the training process is an  NVIDIA Quadro T2000 Max-Q with 1024 CUDA cores of a 1.57 GHz clock rate and 4 GB GDDR5 memory.

The Adam solver \cite{Kingma} was used to update the weights of the two trained operators with a learning rate of 0.001 and a moving average of 0.9. The model was trained for 34 epochs with a batch size of 4 samples. The total time required for the training of the model was approximately 100 hours.

\begin{figure*}[t!]
\centering
\includegraphics[width = \textwidth]{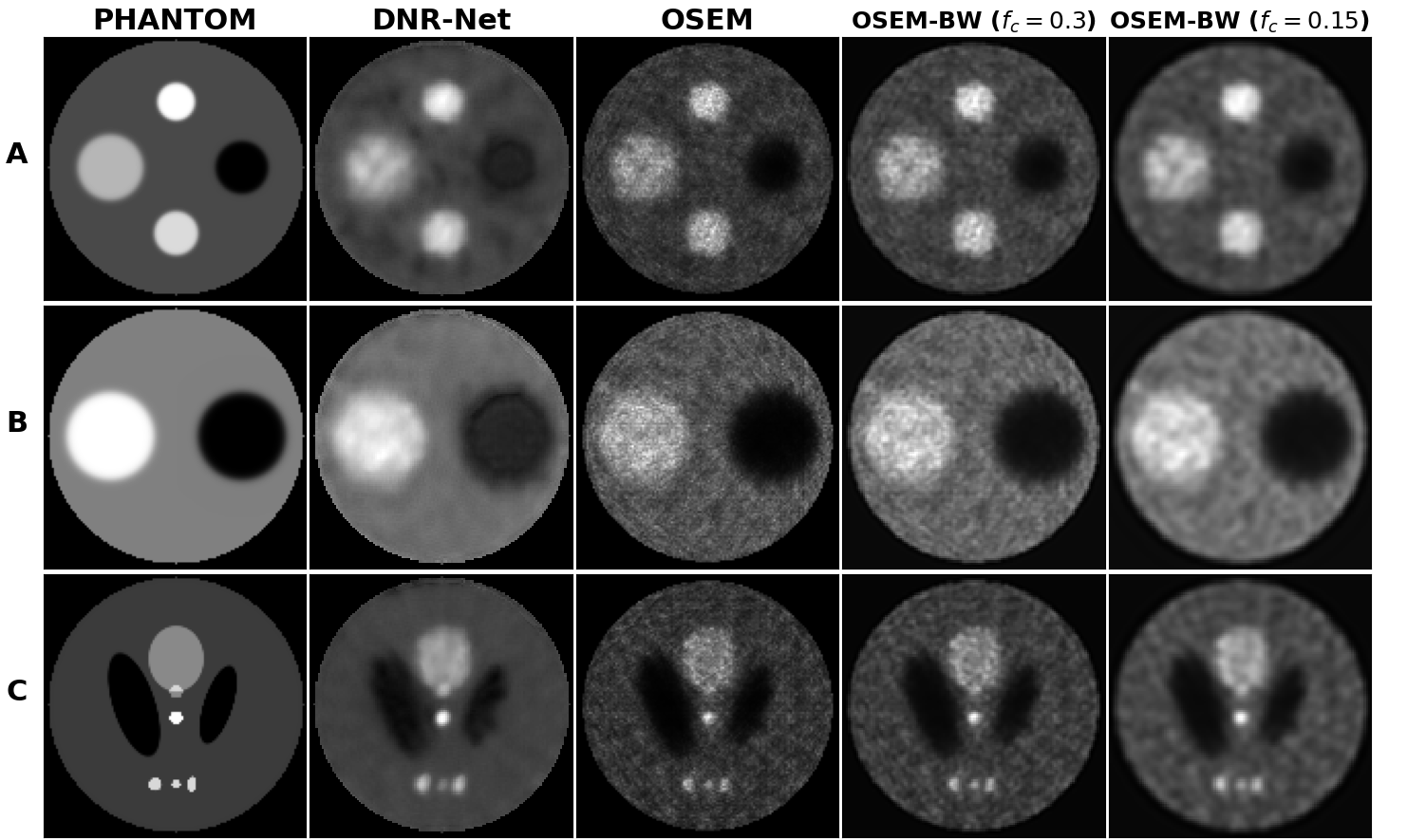}
\caption{The three numerical phantoms used in this study and their reconstructed images were obtained with OSEM and DNR-Net. In all cases, reconstructions were obtained from noisy sinograms containing 24 equally distributed projections. OSEM reconstructions post-filtered with a third-order Butterworth filter of $f_c=0.3$ and $f_c=0.15$ cut-off frequency are also shown. Metrics comparing the quality of the OSEM and DNR-Net images are provided in Table \ref{tab:metrics}.}
\label{Fig:recs}
\end{figure*}

\section{Experiments with Numerical Phantoms}

\subsection{Phantom Simulations and Image Reconstruction}
Three case studies with numerical phantoms were conducted to evaluate the reconstruction capacity of the proposed method. The three numerical phantoms are shown in Fig. \ref{Fig:recs}. Phantom A consists of one coldspot and three hotspots exhibiting different sizes and contrast, embedded in a uniform background. Phantom B features a hotspot and coldspot of the same size embedded in a uniform background of increased activity compared to phantom A. Finally, phantom C is the well-known Shepp Logan phantom having six circular hotspots of different radius and contrast and three rotated ellipsoidal coldspots embedded in a uniform background. Noisy sinograms of 24 projections were generated from the three numerical phantoms following the procedure described in \ref{Sec:Training}. 

All three numerical phantoms used for evaluating the reconstruction capacity of DNR-Net were not included in the training set. The Ordered Subset Expectation Maximization (OSEM) method  \cite{Hudson} was used in this study to provide reference images for comparison. OSEM reconstructions were performed using eight iterations and four subsets. The OSEM images were further post-filtered with a $3^{rd}$ order Butterworth filter of $0.3$ and $0.15$ cycles per pixel cut-off frequency.

\subsection{Image Quality Metrics}
 The quality of the reconstructed images obtained by the two methods is evaluated using the  Structural  Similarity  Index  (SSIM)\cite{Wang}  and the Contrast-to-Noise Ratio (CNR). 
 
 SSIM providing a comparative measure of  luminance, contrast, and structure  is calculated for the reconstructed images $F_r$ using the ground truth images $F_0$  by:
\begin{equation}
    SSIM(F_r, F_0) = \frac{(2\mu_r\mu_0 +C_1)(2\sigma_{r0} +C_2)}{(\mu^2_r + \mu^2_0 +C_1)(\sigma^2_r + \sigma^2_0 +C_2)}
\end{equation}
where $\mu_r$, $\mu_0$, $\sigma_r$, $\sigma_0$ and $\sigma_{r0}$ are the local means, standard deviations and co-variance  of the images $F_r$ and $F_0$, respectively, and $C_1$, $C_2$ are constants.

CNR quantifying the detectability of a Region Of Interest $F_{ROI}$ in a noisy  image is given by:
\begin{equation}
    CNR(F_{ROI}) = \frac{|\mu_{ROI} - \mu_B|}{\sigma_B}
\end{equation}
where $\mu_{ROI}$, $\mu_B$ are the mean activity values of  the ROI and background, respectively, and $\sigma_B$ is the standard deviation of the background noise.

\subsection{Results}
The reconstructed images obtained for the three numerical phantoms with DNR-Net and OSEM are visually compared in Fig. \ref{Fig:recs}. Visually, the reconstruction results demonstrate the capacity of DNR-Net to provide images adequately resolving the three numerical phantoms. DNR-Net accurately reconstructed the activities and geometries of the three numerical phantoms, with the results being close to the ground-truth images. In addition, it can be seen that DNR-Net produced smooth representations of the hotspots and coldspots in the images, which exhibit well-preserved edges and lower levels of background noise compared to OSEM images.   

\begin{table}[t!]
\caption{SSIM and CNR scores comparing the  quality of DNR-Net and OSEM images.}
\centering
\renewcommand{\arraystretch}{1.7}
\label{tab:metrics}
\label{table}
\begin{tabularx}{0.48\textwidth}{>{\raggedright\arraybackslash}p{25pt} >{\centering\arraybackslash}p{15pt} >{\centering\arraybackslash}X >{\centering\arraybackslash}X >{\centering\arraybackslash}X >{\centering\arraybackslash}X}
\hline\hline
Phantom & Metric & DNR-Net & OSEM & OSEM $_{f_c=0.3}$ & OSEM $_{f_c=0.15}$  \\
\hline
A & SSIM & 0.83 & 0.50 & 0.58 & 0.67  \\
  & CNR & 6.0 & 3.8 & 4.4 & 4.9  \\
\hline
B & SSIM & 0.79 & 0.43 & 0.51 & 0.61 \\
  & CNR & 4.4 & 3.0  & 3.4 & 3.8  \\
\hline
C & SSIM & 0.78 & 0.44 & 0.51 & 0.57  \\
  & CNR & 8.4 & 3.9 & 4.6 & 5.0 \\
\hline\hline
\end{tabularx}
\end{table}

SSIM and CNR scores providing a quantitative comparison between the reconstructions by the two methods are given in  Table \ref{tab:metrics}. Overall, the visual observations reported in the previous paragraph are validated by the quantitative results.  As expected, the  OSEM images obtained without post-filtering exhibit the lower CNR and SSIM scores. The scores indicate an improvement in the quality of  OSEM images with the application of the Butterworth filter. The cut-off frequency of 0.15 cycles per pixel led to the best image quality for the  OSEM reconstructions. Compared to those reconstructions obtained with 0.15 cycles per pixel cut-off frequency, DNR-Net produced  higher  SSIM and CNR images for all three reconstruction cases. For the case of Shepp-Logan phantom having the most complex phantom geometry presented here, DNR-Net scored  60$\%$ higher CNR and 37$\%$ higher SSIM compared to the best scores obtained with OSEM.

Overall, the quantitative results indicate that DNR-Net guided by the system matrix of the tomographic problem can provide reconstructions of sufficiently good quality in a general case of a simulated sinogram not used in the training process. Furthermore, without incorporating an explicit regularization function, DNR-Net is able to approximate it through the training process and provide smooth reconstructed images.

\section{Conclusions and Future Research Directions}

This work presented a hybrid data-driven reconstruction algorithm for emission tomography inspired by Newton's method called DNR-Net. The proposed method led to reconstructed images of improved image quality, as quantified through quantitative measures (CNR, SSIM), outperforming the OSEM method in three experiments with simulated sinograms. 

 DNR-Net architecture, configured with six blocks in this work, imitating six sequential steps of Newton's method, was able to approximate Newton's descent direction and provide numerically stable reconstruction results.
 This number of blocks in the DNR-Net configuration is a hyperparameter arbitrarily chosen. In future work, we plan to use parallel computing allied to metaheuristics for optimizing this hyperparameter choice. This way, multiple training could be performed in parallel to generate a population of solutions. 
 
 After this first demonstration of DNR-Net with simulated data, further experimentation is needed for evaluating its reconstruction capacity with real SPECT sinograms.

\section*{Acknowledgements}
The authors would like to thank the NVIDIA AI Technology Center Luxembourg for their positive comments and careful review of the current study.

\def\bibsection{\section*{\refname}}

\end{document}